\documentclass{llncs} %

\usepackage{float}
\usepackage{graphicx}
\usepackage{wrapfig}

\usepackage[labelfont=bf]{caption}
\usepackage{color}
\usepackage[colorlinks=true,linkcolor=blue,citecolor=blue]{hyperref}

\usepackage{color}
\usepackage{amsmath} 
\usepackage{mathtools} 
\usepackage{amsfonts} 
\usepackage{amssymb}
\usepackage[normalem]{ulem}

\usepackage{array}
\usepackage{booktabs}
\usepackage{rotating}
\usepackage{multirow}
\usepackage{multicol}
\usepackage{lscape}
\usepackage{subfig}
\usepackage{comment}
\usepackage[export]{adjustbox}

\usepackage{algorithm,algorithmicx,algpseudocode}

\usepackage{color}

\usepackage[colorlinks=true,linkcolor=blue,citecolor=blue]{hyperref}

\usepackage{tikz,xcolor,hyperref}

\definecolor{lime}{HTML}{A6CE39}
\DeclareRobustCommand{\orcidicon}{
	\begin{tikzpicture}
	\draw[lime, fill=lime] (0,0) 
	circle [radius=0.16] 
	node[white] {{\fontfamily{qag}\selectfont \tiny ID}};
	\draw[white, fill=white] (-0.0625,0.095) 
	circle [radius=0.007];
	\end{tikzpicture}
	\hspace{-2mm}
}

\foreach \x in {A, ..., Z}{\expandafter\xdef\csname orcid\x\endcsname{\noexpand\href{https://orcid.org/\csname orcidauthor\x\endcsname}
			{\noexpand\orcidicon}}
}
			
\usepackage{color}

\definecolor{darkgreen}{rgb}{0.53, 0.66, 0.42}

\begin{document}

\title{StairwayGraphNet for Inter- and Intra-modality Multi-resolution Brain Graph Alignment and Synthesis}

\titlerunning{StairwayGraphNet for Inter- and Intra-modality Multi-resolution}  

\author{Islem Mhiri\orcidC{}\index{Mhiri, Islem}\inst{1, 2}  \and Mohamed Ali Mahjoub\orcidD{} \index{Mahjoub, Mohamed Ali}\inst{1} \and Islem Rekik\orcidA{} \index{Rekik, Islem}\inst{2}\thanks{ {corresponding author: \url{irekik@itu.edu.tr}. This project has been funded by the TUBITAK 2232 Fellowship (Project No:118C288).}} }

\authorrunning{I Mhiri et al.}

\institute{$^{1}$ BASIRA Lab, Faculty of Computer and Informatics, Istanbul Technical University, Istanbul, Turkey (\url{http://basira-lab.com/}) \\ $^{2}$ Universit\'e de Sousse, Ecole Nationale d'Ing\'enieurs de Sousse, LATIS- Laboratory of Advanced Technology and Intelligent Systems, 4023, Sousse, Tunisie}

\maketitle              

\begin{abstract}

Synthesizing multimodality medical data provides complementary knowledge and helps doctors make precise clinical decisions. Although promising, existing multimodal brain graph synthesis frameworks have several limitations. First, they mainly tackle only one problem (intra- or inter-modality), limiting their generalizability to synthesizing inter- and intra-modality simultaneously. Second, while few techniques work on super-resolving low-resolution brain graphs within a single modality (i.e., intra), inter-modality graph super-resolution remains unexplored though this would avoid the need for costly data collection and processing.
More importantly, both target and source domains might have different distributions, which causes a domain fracture between them. To fill these gaps, we propose a multi-resolution StairwayGraphNet (SG-Net) framework to jointly infer a target graph modality based on a given modality and super-resolve brain graphs in both inter and intra domains. Our SG-Net is grounded in three main contributions: (i) predicting a target graph from a source one based on a novel graph generative adversarial network in both inter (e.g., morphological-functional) and intra (e.g., functional-functional) domains, (ii) generating high-resolution brain graphs without resorting to the time consuming and expensive MRI processing steps, and (iii) enforcing the source distribution to match that of the ground truth graphs using an inter-modality aligner to relax the loss function to optimize. Moreover, we design a new Ground Truth-Preserving loss function to guide both generators in learning the topological structure of ground truth brain graphs more accurately. Our comprehensive experiments on predicting target brain graphs from source graphs using a multi-resolution stairway showed the outperformance of our method in comparison with its variants and state-of-the-art method. SG-Net presents the first work for graph alignment and synthesis across varying modalities and resolutions, which handles graph size, distribution, and structure variations.  Our Python SG-Net code is available on BASIRA GitHub at \url{https://github.com/basiralab/SG-Net}.
\end{abstract}

\section{Introduction}
Multimodal brain imaging has shown tremendous potential for neurological disorder diagnosis, where each imaging modality offers specific information for learning more holistic and informative data representations. Although doctors require many imaging modalities to aid in precise clinical decision, they suffer from restricted medical conditions such as high acquisition cost and processing time \cite{cao2020}. To address these limitations, several deep learning studies have investigated multimodal MRI synthesis \cite{yu2020}. Such methods either synthesize one modality from another (i.e., cross-modality) or map both modalities to a commonly shared domain. Notably, generative adversarial networks (GANs) \cite{singh2021,wang2021} have achieved great success in predicting medical images of different brain image modalities from a given modality. For instance, \cite{liu2020} proposed a joint neuroimage synthesis and representation learning framework with transfer learning for subjective cognitive decline conversion prediction where they imputed missing PET images using MRI scans. Also, \cite{zhan2021} aimed to synthesize a missing MRI modality from multiple modalities using conditional GAN. 
Although significant clinical representations were obtained from the latter studies, more substantial challenges still exist. The brain connectome has a complex non-linear structure that is difficult to capture using linear models \cite{van2019}. Moreover, many approaches do not make effective use of or even fail to handle non-Euclidean structured data (i.e., geometric data), such as graphs and manifolds \cite{bronstein2017}. Therefore, a deep learning model that retains graph-based data representation topology provides a relevant research direction to be explored. 

Recently, geometric deep learning techniques have shown  great potential in learning the deep graph-structure. Particularly, deep graph convolutional networks (GCNs) have imbued the field of network neuroscience research through various tasks such as studying the mapping between human connectome and disease outcome \cite{bessadok2021}. Recent landmark studies have relied on using GCN to predict a target brain graph from a source brain graph. 
For instance, \cite{bessadok2020} introduced MultiGraphGAN architecture, which predicts multiple brain graphs from a single brain graph while preserving the topological structure of each target predicted graph. Moreover, \cite{zhang2020b} defined a multi-GCN-based generative adversarial network (MGCN-GAN) to synthesize individual structural connectome from a functional connectome. However, all these works can only transfer brain graphs from \emph{one modality} to another while preserving the same resolution, limiting their generezability to \emph{cross-modality} brain graph synthesis at \emph{different resolutions} (i.e., node size). Therefore, using multi-resolution graphs (e.g., super-resolution) remains a significant challenge in designing generalizable and scalable brain graph synthesis models. \cite{isallari2020} circumvented this issue by designing a graph neural network for super-resolving low-resolution (LR) functional brain connectomes from a single modality. However, super-resolving brain graphs across modalities (i.e., inter) is strikingly lacking. Furthermore, most previous works still face domain fracture problems resulting in the difference in distribution between the source and target domains. Remarkably, domain alignment remains mostly scarce in brain graph synthesis tasks \cite{bessadok2021}. While yielding outstanding performance, all the aforementioned methods have tackled only one problem (inter \emph{or} intra-modality), which hinders their generalizability to synthesizing inter and intra-modalities jointly. 

 
To address the challenges above and motivated by the recent development of graph neural network-based solutions, we propose a multi-resolution StairwayGraphNet (SG-Net) method to jointly predict and super-resolve a target graph modality based on a given modality in both inter- and intra-domains. To do so, prior to the prediction blocks, we propose an inter-modality aligner network based on adversarially regularized variational graph autoencoder (ARVGA) \cite{pan2018} to align the training graphs of the source modality to that of the target one. Second, given the aligned source graphs, we design an inter-modality super-resolution graph GAN (gGAN) to map the aligned source graph from one modality (e.g., morphological) to the target modality (e.g., functional). Note that the alignment step facilitates the training of our super-resolution gGAN since both source and target domains have been aligned by the inter-modality aligner network (i.e., shared mode). To capture the complex relationship in both direct and indirect brain connections, we design the super-resolution generator and discriminator of our inter-modality GAN using edge-based GCNs \cite{simonovsky2017}. Then, we synthesize high-resolution (HR) functional brain graphs from LR functional graphs using GCN-based intra-modality GAN. 
Besides, to resolve the inherent instability of GANs, we propose a novel ground-truth-preserving (GT-P) loss function to enforce our multi-resolution generators to effectively learn the ground-truth brain graph.

The main contributions of our work are four-fold. \textbf{On a methodological level,} StairwayGraphNet presents the first work for graph alignment and synthesis across varying modalities and resolutions, which can also be leveraged for boosting neurological disorder diagnosis. \textbf{On a clinical level,} learning multi-resolution brain connectivity synthesis can provide comprehensive brain maps that capture multimodal relationships (functional, structural, etc.) between brain regions, thereby charting brain dysconnectivity patterns in disordered populations \cite{van2019}. \textbf{On a computational level,} our method generates HR graphs without resorting to any computational MRI processing step such as registration and parcellation. \textbf{On a generic level,} our framework is a generic method as it can be applied to predict brain graphs derived from any neuroimaging modalities with different resolutions and complex nonlinear distributions.

\begin{figure}[t!]
\centering
\includegraphics[width=12cm]{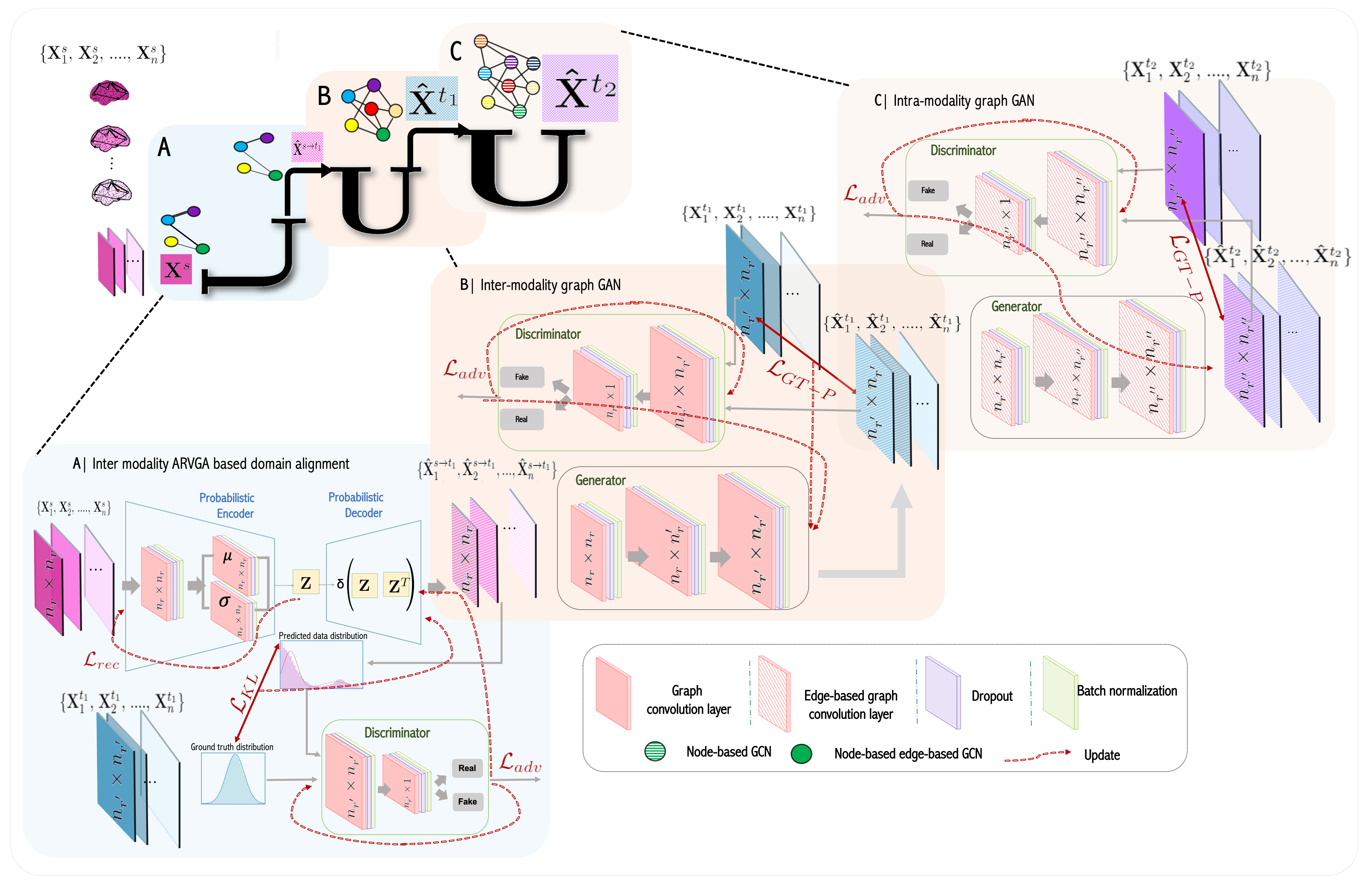}
\vspace{-10pt}
\caption{\emph{ Illustration of the proposed inter and intra-modality multi-resolution brain graph alignment and synthesis using SG-Net.} \textbf{A$\mid$ Graph-based inter-modality aligner.} We aim to align the training graphs of the source modality $\mathbf{X}^{s}$ to that of the target one $\mathbf{X}^{t_{1}}$. Thus, we design an ARVGA model with KL-divergence to bridge the gap between the distributions of the source and target graphs. \textbf{B$\mid$ Adversarial inter-modality graph GAN.} Next, we propose an inter-modality multi-resolution graph GAN to transform the aligned source brain graph $\mathbf{\hat{X}}^{s \rightarrow t1}$ (e.g., morphological) into the target graph (e.g., functional) with different structural and topological properties. \textbf{C$\mid$ Adversarial intra-modality super-resolution graph GAN.} Then, we super-resolve the predicted functional graphs $\mathbf{\hat{X}}^{t_{1}}$ into a HR brain graphs $\mathbf{\hat{X}}^{t_{2}}$ using an intra-modality super-resolution GAN architecture. The aligner and both generators are trained in an end-to-end manner by optimizing a novel Ground Truth-Preserving (GT-P) loss function which guides them in learning the topology of the ground truth brain graphs more effectively.} 
\label{fig:1}
\end{figure}
\section{Proposed method}
This section presents the key steps of our stairway graph alignment, prediction, and super-resolution (SG-Net) framework. 
\textbf{Fig.}~\ref{fig:1} presents an overview for the proposed framework in three major steps, which are detailed below.

\textbf{Problem Definition.} A brain graph can be represented as $\mathbf{G} (\mathbf{V}, \mathbf{E})$, where each node in $\mathbf{V}$ denotes a brain region of interest and each edge in $\mathbf{E}$ connecting two ROIs $k$ and $l$ denotes the strength of their connectivity. Each training subject $i$ is represented by three brain graphs $\{\mathbf{G}^s_i (\mathbf{V}^s_i, \mathbf{E}^s_i), \mathbf{G}_i^{t_{1}} (\mathbf{V}_i^{t_{1}}, \mathbf{E}_i^{t_{1}}), \mathbf{G}_i^{t_{2}} (\mathbf{V}_i^{t_{2}}, \mathbf{E}_i^{t_{2}})\}$, where $\mathbf{G}^{s}$ is the source brain graph (morphological brain network) with $n_r$ nodes, $\mathbf{G}^{t_1}$ is the first target brain graph (LR functional brain network) with $n_{r^{'}}$ nodes and $\mathbf{G}^{t_2}$ is the second target brain graph (HR functional brain network) with $n_{r^{''}}$ nodes where $n_r \ne n_{r^{'}} \ne n_{r^{''}}$. 

\textbf{A- Graph-based inter-modality aligner block}. Existing cross-domain frameworks are mainly tailored for images. However, since graphs are in the non-Euclidean space, the cross-domain prediction for graphs is exceedingly challenging. Hence, we propose a graph-based inter-modality aligner framework that enforces the predictions of the source graphs model to have the same distribution as the ground truth graphs. 
 Specifically, given a set of $n$ training source brain networks (e.g., morphological connectomes) $\mathbf{X}^s_{tr}$ and a set of $n$ training ground-truth brain networks (e.g., functional connectomes) $\mathbf{X}^{t_1}_{tr}$, for each subject $i$, our aligner takes $\mathbf{X}^s_i$ as input and outputs $\mathbf{\hat{X}}_i^{s \rightarrow {t_1}}$ which shares the same distribution of $\mathbf{X}^{t_1}_i$ (\textbf{Fig.}~\ref{fig:1}-A). Drawing inspiration from ARVGA \cite{pan2018}, our framework is composed of a variational autoencoder $A_{align}$ and a discriminator $D_{align}$. The $A_{align}$ comprises a probabilistic encoder that encodes an input as a distribution over the latent space instead of encoding an input as a single point. Indeed, the ARVGA produces a vector of mean $\mu$ and standard deviation $\sigma$, which provides more continuity in the latent space than the conventional autoencoder. This continuity enables the probabilistic decoder not only to reproduce an input vector but also to generate new data from the latent space \cite{zemouri2020}. To provide a more effective embedding, we propose an adversarial learning scheme using a discriminator $D_{align}$ that \textbf{enforces} the latent representation to match that of the prior distribution of the ground-truth. Specifically, our encoder comprises three edge-based GCN layers adjusted by adding batch normalization and dropout to each layer's output (\textbf{Fig.}~\ref{fig:1}-A). Indeed, batch normalization efficiently accelerates the network training through a fast convergence of the loss function, and dropout reduces the possibility of overfitting. Therefore, these two operations help optimize and simplify the network training.
To refine our inter-modality aligner, we propose an alignment loss function:
$\mathcal{L}_{align}= \lambda_{adv} \mathcal{L}_{adv}+\lambda_{rec} \mathcal{L}_{rec}+\lambda_{KL} \mathcal{L}_{KL}$, where $\mathcal{L}_{adv}$ \cite{goodfellow2014} is the adversarial loss quantifying the difference between the generated and ground truth target graphs since both $A_{align}$ and $D_{align}$ are iteratively optimized. $\mathcal{L}_{rec} = \left\|\mathbf{X}^{s}-\mathbf{\hat{X}}^{s \rightarrow t_1}\right\|^{2}$ denotes the reconstruction loss that tends to enhance the encoding-decoding scheme, where $\mathbf{X}^{s}$ is the source brain graph and $\mathbf{\hat{X}}^{s \rightarrow t_1}$ is the aligned source graphs. $\mathcal{L}_{KL}$ is Kulback-Leibler (KL) divergence loss which acts as a regularization term. $\mathcal{L}_{KL}$, also known as the relative entropy, is an asymmetric measure that quantifies the difference between two probability distributions.
Thereby, we use $\mathcal{L}_{KL}$ to minimize the discrepancy between ground-truth and aligned source brain graph distributions.
$\mathcal{L}_{KL}= \sum_{i=1}^{n} KL\left(q_{i} \| p_{i}\right)$, where the KL divergence for subject $i$ is defined as: $ KL\left(q_{i} \| p_{i}\right)=\int_{-\infty}^{+\infty} q_{i}(x) \log \frac{q_{i}(x)}{p_{i}(x)} d x$ where $q$ is the true distribution (ground truth) and $p$ is the aligned.

\textbf{B- Adversarial inter-modality graph generator block.} Following the inter-modality alignment step, we design an inter-modality generator that handles shifts in graph resolution (i.e., node size variation) coupled with an adversarial discriminator.

\emph{i- Inter-modality multi-resolution brain graph generator.} Inspired by the dynamic edge convolution proposed in and \cite{simonovsky2017} the U-net architecture \cite{ronneberger2015} with skip connections, our inter-modality graph generator $G_{inter}$ is composed of three edge-based GCN layers regularized by batch normalization and dropout for each layer's output (\textbf{Fig.}~\ref{fig:1}-B). $G_{inter}$ takes as input the aligned source graphs to the target distribution $\mathbf{\hat{X}}^{s \rightarrow t_1}$ of size $n_r \times n_r$ and outputs the predicted target brain graphs $\mathbf{\hat{X}}^{t_1}_{i}$ of size $n_{r^{'}} \times n_{r^{'}}$, where $n_{r} \ne n_{r^{'}}$ . Particularly, owing to \cite{simonovsky2017}, each edge-based GCN layer includes a unique dynamic filter that outputs edge-specific weight matrix dictating the information flow between nodes $k$ and $l$ to learn a comprehensive vector representation for each node ($\mathbf{z}{_{k}^{h}} \in \mathbb{R}^{1 \times d_h}$ is the embedding of node $k$ in layer $h$ where $d_h$ denotes the output dimension of the corresponding layer). Next, to learn our inter-modality multi-resolution mapping, we define a mapping function $\mathcal{T}_{r}: \mathbb{R}^{n_r \times d_h} \mapsto \mathbb{R}^{d_h \times d_h}$ that takes as input the embedded matrix of the whole graph $\mathbf{Z}^{h}$ in layer $h$ of size ${n_r \times d_h}$ and outputs the generated target graph of size $d_h \times d_h$. We formulate $\mathcal{T}_{r}$ as follows: $\mathcal{T}_{r} = ({\mathbf{Z}^{h}})^T \mathbf{Z}^{h}$.
As such, shifting resolution is only defined by fixing the desired target graph resolution $d_{h}$. In our case, we set $d_h$ of the latest layer in the generator to $n_{r^{'}}$ to output the predicted target brain graph $\mathbf{\hat{X}}^{t_1}$ of size $n_{r^{'}} \times n_{r^{'}}$ (\textbf{Fig.}~\ref{fig:1}-B).

\emph{ii- Inter-modality graph discriminator based on adversarial training.} Our inter-modality generator $G_{inter}$ is trained adversarially against a discriminator network $D_{inter}$ (\textbf{Fig.}~\ref{fig:1}-B). To discriminate between the predicted and ground truth target graph data, we design a two-layer graph neural network \cite{simonovsky2017}. Our discriminator $D_{inter}$ takes as input the real brain graph $\mathbf{X}^{t_1}_i$ and the generator's output $\mathbf{\hat{X}}^{t_1}_i$, and outputs a value between $0$ and $1$, measuring the generator's output's realness. To boost our discriminator's performances, we adopt the adversarial loss function to maximize the discriminator's output value for the $\mathbf{X}^{t_1}_i$ and minimize it for $\mathbf{\hat{X}}^{t_1}_i$.

\textbf{C- Adversarial intra-modality graph generator block.} To avoid time-consuming image processing pipelines, we aim to predict intra-modality brain graphs at higher resolutions. Specifically, we design an intra-modality GAN similar to inter-modality GAN, yet without alignment and using node-based GCN layer \cite{kipf2016} instead of edge-based GCN (\textbf{Fig.}~\ref{fig:1}-C). In fact, we used edge-based GCN for the inter-domain generation since simultaneously transferring one modality to another and super-resolving brain graphs are difficult tasks requiring more robust GCN layers. But, as we super-resolve up the network \emph{stairs} (\textbf{Fig.}~\ref{fig:1}), edge-based GCN becomes highly time consuming and RAM-devouring. Thus, we use node-based GCN for super-resolving brain graphs from the same modality, which is less computationally expensive.

\textbf{D- Ground truth-Preserving loss function.}
Conventionally, GAN generators are optimized based on the response of their corresponding discriminators. However, within a few training epochs, the discriminator can easily distinguish real graphs from predicted graphs, and the adversarial loss would be close to 0. In such a case, the generator cannot provide satisfactory results. To overcome this dilemma, we need to enforce the generator-discriminator synchronous learning during the training process.
Thus, we propose a new ground truth-preserving (GT-P) loss for both intra and inter-modality prediction blocks (\textbf{Fig.}~\ref{fig:1}-B and C). Our loss is composed of four sub-losses: adversarial loss \cite{goodfellow2014}, $L1$ loss \cite{gurler2020}, Pearson correlation coefficient (PCC) loss \cite{zhang2020b}, and topological loss. We define our GT-P loss function as follows:
$\mathcal{L}_{\text {GT-P}}= \lambda_1 \mathcal{L}_{adv}+\lambda_2 \mathcal{L}_{L 1}+\lambda_3 \mathcal{L}_{PCC}+\lambda_4 \mathcal{L}_{top}$.
To improve the predicted target brain graph quality, we propose to add an $L1$ loss term minimizing the distance between each predicted subject $\mathbf{\hat{X}}^{t}$ and its related ground truth $\mathbf{X}^{t}$. 
Even robust to outliers, the $L1$ loss only focuses on the element-wise similarity in edge weights between the predicted and real brain graphs and ignores the overall correlation between both graphs. Hence, we include the Pearson correlation coefficient (PCC) in our loss which measures the overall correlation between the predicted and real brain graphs. Since higher PCC indicates a higher correlation between the ground-truth and the predicted graphs, we propose to minimize the PCC loss function as follows: $\mathcal{L}_{PCC}= 1 - PCC$.

Furthermore, each brain graph has its unique topology, we introduce a topological loss function that guides the generator to maintain the nodes' topological profiles while learning the global graph structure. To do so, we define the $L1$ loss between the real and predicted eigenvector centralities (ECs) capturing the centralities of a node's neighbors. Hence, we define our topology loss as $\mathcal{L}_{top}=\left\|\mathbf{c}^{t}-\mathbf{\hat{c}}^{t}\right\|_{1}$, where $\mathbf{\hat{c}}^{t}$ denotes the EC vector of the predicted brain graph and $\mathbf{c}^{t}$ is the EC vector of the real one. 
\section{Results and Discussion}
\textbf{Evaluation dataset and parameters.} We used three-fold cross-validation to evaluate our SG-Net framework on 150 subjects from the Southwest University Longitudinal Imaging Multimodal (SLIM) public dataset\footnote{\url{http://fcon\_1000.projects.nitrc.org/}} where each subject has T1-w, T2-w MRI and resting-state fMRI (rsfMRI) scans. Our model is implemented using the PyTorch-Geometric library \cite{fey2019}. Following several preprocessing steps \cite{fischl2012}, a $35 \times 35$ cortical morphological network was generated from structural T1-w MR for each subject denoted as $\mathbf{X}^{s}$. For the resting-state functional MRI images, two separate brain networks with $160 \times 160$ (LR) denoted as $\mathbf{X}^{t_1}$and $268 \times 268$ (HR) denoted as $\mathbf{X}^{t_2}$ were produced for each subject using two group-wise whole-brain parcellation approaches proposed in \cite{dosenbach2010} and \cite{shen2013}, respectively. For the aligner hyperparameters, we set $\lambda_{adv} = 1 $, $\lambda_{rec} = 0.1$ and $\lambda_{KL} = 0.001$. Also, we set both generators hyperparameters as follows: $\lambda_1 = 1$, $\lambda_2 = 1$, $\lambda_3 = 0.1$, and $\lambda_4 = 2$. Moreover, we chose AdamW \cite{loshchilov2018} as our default optimizer and set the learning rate at $0.025$ for the $A_{align}$ and the generators networks, and $0.01$ for all the discriminators. Finally, we trained our model for 400 epochs using a single Tesla V100 GPU (NVIDIA GeForce GTX TITAN with 32 GB memory).

\begin{table*}[h]
\caption{\emph{Prediction results using different evaluation metrics.} Evaluation of alignment and prediction brain graph synthesis by our SG-Net against seven comparison methods. CC: closeness centrality, BC: betweenness centrality and EC: eigenvector centrality. w/o: without. $\star$: SG-Net significantly outperformed all benchmark methods using two-tailed paired t-test ($p < 0.05$).}
\begin{tabular}{c c c c c c}
\hline
{ Methods}                     & { MAE}    & { MAE (BC)} & { MAE (CC)} & { MAE (EC)} \\ \hline
{ SG-Net w/o alignment  }  & { 0.382} & { 0.035} & { 0.76} & { 0.0171} \\ 
{ Statistical alignment based-SG-Net} & { 0.345} & { 0.031} & { 0.42} & { 0.0166}  \\ 
{ SG-Net using VGAE \cite{pan2018} based-alignment}     & { 0.191} & { 0.012} & { 0.21} & { 0.0148} \\ 
{ SG-Net using ARGA \cite{pan2018} based-alignment}   & { 0.136} & {  0.010} & { 0.182}  & { 0.0142} \\ \hline \hline { GSR-Net \cite{isallari2020} }     & { 0.253} & {\underline{ 0.0083}} & { \underline{0.1447}} & { \textbf{0.0114}}\\\hline \hline
{ SG-Net w/o PCC}      & { 0.115} & { 0.0087} & { 0.1448} & { 0.0115}
\\ 
{ SG-Net w/o topology}      & { \underline{0.102}} & { 0.0094} & { 0.1473} & {0.0120}\\
{ \textbf{SG-Net}}      & {  $\textbf{0.097}^{\star}$} & { $\textbf{0.0073}^{\star}$} & { $\textbf{0.1439}^{\star}$} & { $\textbf{0.0114}^{\star}$} \\
\hline
\end{tabular}

\label{tab:1}
\end{table*}

\begin{table*}[h]
\caption{\emph{Inter-modality prediction results using different evaluation metrics (part B).} Evaluation of alignment and prediction inter-modality brain graph synthesis by our SG-Net against seven comparison methods. CC: closeness centrality, BC: betweenness centrality and EC: eigenvector centrality. . w/o: without. $\star$: SG-Net significantly outperformed all benchmark methods using two-tailed paired t-test ($p < 0.05$).}
\begin{tabular}{c c c c c c}
\hline
{ Methods}                     & { MAE}    & { MAE (BC)} & { MAE (CC)} & { MAE (EC)} \\ \hline
{ SG-Net w/o alignment  }  & { 0.376} & { 0.041} & { 0.715} & { 0.0243} \\ 
{ Statistical alignment based-SG-Net} & { 0.351} & { 0.028} & { 0.712} & { 0.0169}  \\ 
{ SG-Net using VGAE based-alignment}     & { 0.181} & { 0.015} & { 0.48} & { 0.0151} \\ 
{ SG-Net using ARGA  based-alignment}   & { 0.164} & {  0.013} & { 0.37}  & { 0.0147} \\
{ SG-Net w/o PCC}      & { 0.110} & { \underline{0.0094}} & { \underline{0.152}} & { \underline{0.0132}}
\\ 
{ SG-Net w/o topology}      & { \underline{0.106}} & { 0.001} & { 0.23} & {0.0136}\\
{ \textbf{SG-Net}}      & {  $\textbf{0.103}^{\star}$} & { $\textbf{0.0091}^{\star}$} & { $\textbf{0.1514}^{\star}$} & { $\textbf{0.0127}^{\star}$} \\
\hline
\end{tabular}

\label{tab:2}
\end{table*}

\begin{figure}[htpb]
\centering
\includegraphics[width=12cm]{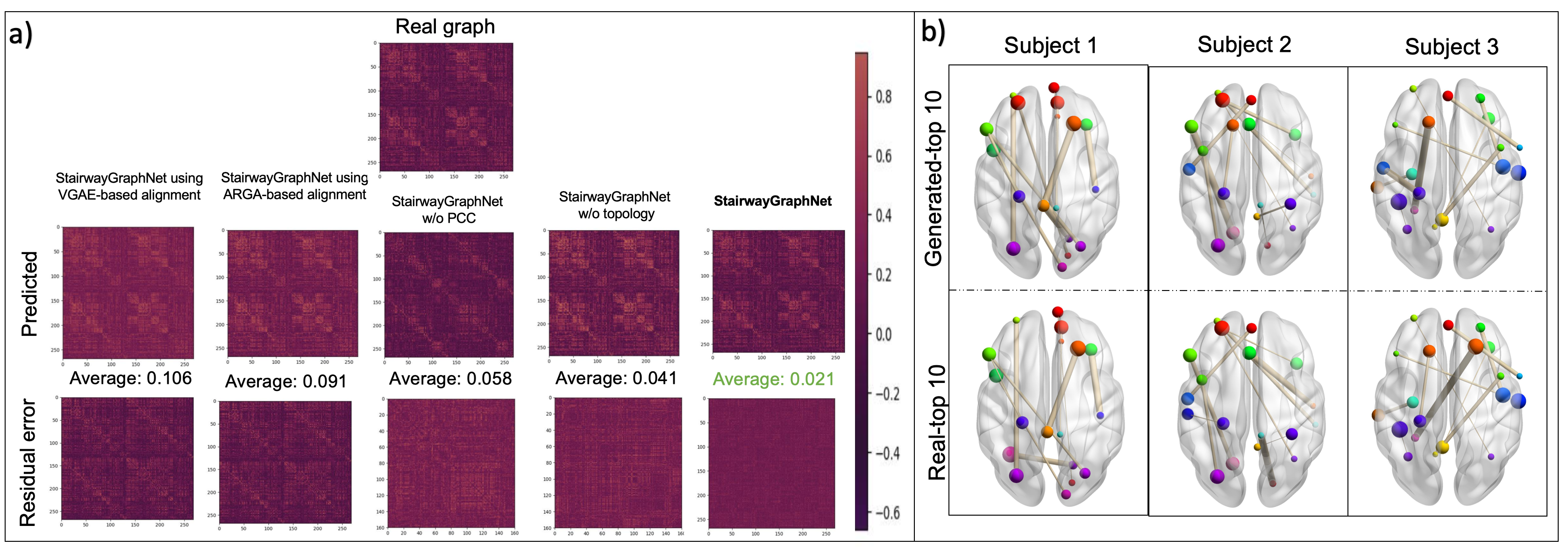}
\caption{\emph{Visual comparison between the real and the predicted inter-modality brain graph distributions (part A).} Evaluation of the distribution alignment between the ground truth to the predicted inter-modality brain graphs by SG-Net against three comparison methods using KL divergence.} 
\label{fig:3}
\end{figure}

\textbf{Evaluation and comparison methods.} \emph{Brain graph alignment.} To evaluate the performance of our aligner, we carried out four major comparisons. As shown in \textbf{Table}~\ref{tab:1} and \textbf{Table}~\ref{tab:2} in both inter and intra-domains, SG-Net w/o alignment method scored the highest MAE between the real and predicted brain graphs.
This demonstrates that the domain alignment improves the quality of the generated
graphs. Moreover, according to \textbf{Table}~\ref{tab:2} and \textbf{Fig.}~\ref{fig:3}, when using a complex non-linear distribution (our source morphological distribution in \textbf{Fig.}~\ref{fig:1}), the statistical aligner may not align to the target distribution properly. Undeniably, an \emph{adversarial learning-based} aligner can better adapt to any distribution, thereby achieving better results.

\emph{Insights into topological measures.} To prove the fidelity of the predicted brain graphs to the real ones in topology and structure, we tested SG-Net using various topological measures (eigenvector, closeness, and betweenness). We note that the methods using the topological loss (GSR-Net \cite{isallari2020}, SG-Net w/o PCC and SG-Net) achieved better results.
Specifically, our method produced the lowest MAE between the ground truth and
predicted brain graphs in both inter- and intra-domains across all topological measurements ((\textbf{Table}~\ref{tab:1} and \textbf{Table}~\ref{tab:2}), showing that our model can better preserve the topology of the predicted functional connectomes.

\emph{Insights into the proposed GT-P loss function.}
To investigate the efficiency of the proposed GT-P loss, we trained SG-Net with different loss functions. As demonstrated in \textbf{Table}~\ref{tab:1}) and  \textbf{Table}~\ref{tab:2}, our proposed GT-P loss outperforms its ablated versions and state-of-the-art method. In fact, the $L1$ loss focuses only on minimizing the distance between two brain graphs at the local level. Besides, PCC aims to maximize global connectivity patterns between the predicted and real brain graphs. However, both losses overlook the topological properties of graphs. Therefore, the EC is introduced in our topological loss to quantify a node's influence on a network's information flow. The combination of these complementary losses scored the best results while relaxing both intra-and inter-modality graph hypothesis between source and target domains.

\begin{figure}[htpb]
\centering
\includegraphics[width=12cm]{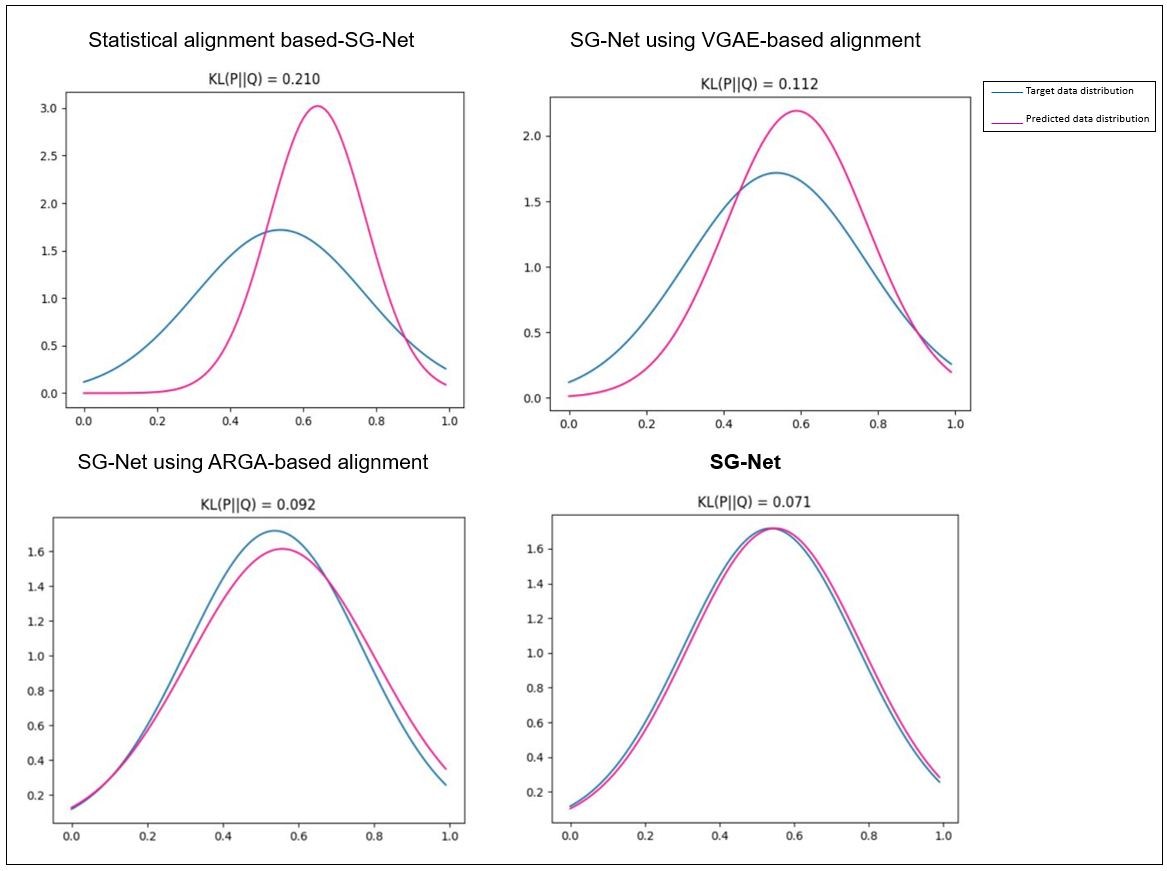}
\caption{\emph{Visual comparison between the real and the predicted inter-modality brain graphs (part B).} Comparing the ground truth to the predicted inter-modality brain graphs by SG-Net and five comparison methods using a representative testing subject. We display the residual matrices computed using the absolute difference between ground truth and predicted connectivity matrices.} 
\label{fig:4}
\end{figure}

\begin{figure}[htpb]
\centering
\includegraphics[width=12cm]{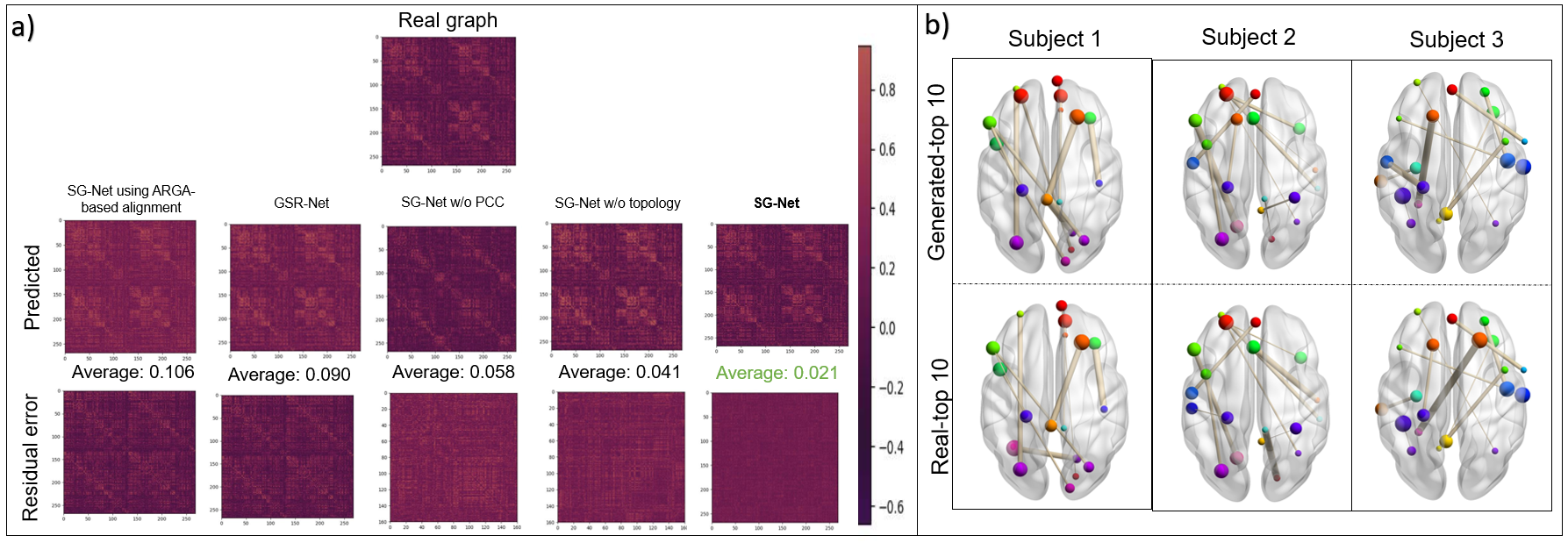}
\caption{\emph{Visual comparison between the real and the predicted target brain graphs.} a) Comparing the ground truth to the predicted brain graphs by SG-Net and five comparison methods using a representative testing subject. We display the residual matrices computed using the absolute difference between ground truth and predicted brain graph connectivity matrices. b) The top 10 strongest connectivities of real and predicted HR functional brain networks of 3 randomly selected testing subjects.} 
\label{fig:2}
\end{figure}

\emph{Reproducibility.} Besides generating realistic HR functional brain graphs, our framework could also capture the delicate variations in connectivity patterns across subjects. Specifically, we display in (\textbf{Fig.}~\ref{fig:2}-a) and  \textbf{Fig.}~\ref{fig:4}, the real, predicted, and residual brain graphs for a representative testing subject using five different methods in both inter and intra-domains. The residual graph is calculated by computing the absolute difference between the real and predicted brain graphs. An average difference value of the residual is represented on top of each residual graph achieving a noticeable reduction by SG-Net.

\emph{Neuro-biomarkers.} \textbf{Fig.}~\ref{fig:2}-b displays the top $10$ strongest connectivities of real and predicted HR functional brain graphs of $3$ randomly selected testing subjects. Since brain connectivity patterns differ from an individual to another \cite{glasser2016}, we note that the top $10$ connectivities are not identical. Yet, our model can accurately predict such variations as well as individual trends in HR functional connectivity based on morphological brain graphs derived from the conventional T1-w MRI. This result further indicates that our method is trustworthy for synthesizing \emph{multimodal} brain dysconnectivity patterns in disordered populations \cite{van2019} from limited neuroimaging resources (only T1-w MRI)


\section{Conclusion}
In this paper, we proposed the first work for inter- and intra-modality multi-resolution brain graph alignment and synthesis, namely StairwayGraphNet, which nicely handles variations in graph distribution, size, and structure. Our method can also be leveraged for developing precision medicine as well as multimodal neurological disorder diagnosis frameworks. The proposed SG-Net outperforms the baseline methods in terms of alignment, prediction, and super-resolution results. SG-Net not only predicts reliable HR functional graphs from morphological ones but also preserves the topology of the target domain. In future work, we will extend our architecture to predict multiple HR modality graphs from a single source one.

\section{Acknowledgments}

This work was funded by generous grants from the European H2020 Marie Sklodowska-Curie action (grant no. 101003403, \url{http://basira-lab.com/normnets/}) to I.R. and the Scientific and Technological Research Council of Turkey to I.R. under the TUBITAK 2232 Fellowship for Outstanding Researchers (no. 118C288, \url{http://basira-lab.com/reprime/}). However, all scientific contributions made in this project are owned and approved solely by the authors.

\section{Supplementary material}

We provide three supplementary items for reproducible and open science:

\begin{enumerate}
	\item A 5-mn YouTube video explaining how our framework works on BASIRA YouTube channel at \url{https://youtu.be/EkXcReAeGdM}.
	\item SG-Net code in Python on GitHub at \url{https://github.com/basiralab/SG-Net}. 
\end{enumerate}

\bibliography{Biblio3}

\begin{thebibliography}{10}

\bibitem{cao2020}
Cao, B., Zhang, H., Wang, N., Gao, X., Shen, D.:
\newblock Auto-gan: self-supervised collaborative learning for medical image
  synthesis.
\newblock Proceedings of the AAAI Conference on Artificial Intelligence (2020)
  10486--10493

\bibitem{yu2020}
Yu, B., Wang, Y., Wang, L., Shen, D., Zhou, L.:
\newblock Medical image synthesis via deep learning.
\newblock Deep Learning in Medical Image Analysis (2020)  23--44

\bibitem{singh2021}
Singh, N.K., Raza, K.:
\newblock Medical image generation using generative adversarial networks: A
  review.
\newblock Health Informatics: A Computational Perspective in Healthcare (2021)
  77--96

\bibitem{wang2021}
Wang, C., Yang, G., Papanastasiou, G., Tsaftaris, S.A., Newby, D.E., Gray, C.,
  Macnaught, G., MacGillivray, T.J.:
\newblock Dicyc: Gan-based deformation invariant cross-domain information
  fusion for medical image synthesis.
\newblock Information Fusion \textbf{67} (2021)  147--160

\bibitem{liu2020}
Liu, Y., Pan, Y., Yang, W., Ning, Z., Yue, L., Liu, M., Shen, D.:
\newblock Joint neuroimage synthesis and representation learning for conversion
  prediction of subjective cognitive decline.
\newblock International Conference on Medical Image Computing and
  Computer-Assisted Intervention (2020)  583--592

\bibitem{zhan2021}
Zhan, B., Li, D., Wang, Y., Ma, Z., Wu, X., Zhou, J., Zhou, L.:
\newblock Lr-cgan: Latent representation based conditional generative
  adversarial network for multi-modality mri synthesis.
\newblock Biomedical Signal Processing and Control \textbf{66} (2021)  102457

\bibitem{van2019}
van~den Heuvel, M.P., Sporns, O.:
\newblock A cross-disorder connectome landscape of brain dysconnectivity.
\newblock Nature reviews neuroscience \textbf{20} (2019)  435--446

\bibitem{bronstein2017}
Bronstein, M.M., Bruna, J., LeCun, Y., Szlam, A., Vandergheynst, P.:
\newblock Geometric deep learning: going beyond euclidean data.
\newblock IEEE Signal Processing Magazine \textbf{34} (2017)  18--42

\bibitem{bessadok2021}
Bessadok, A., Mahjoub, M.A., Rekik, I.:
\newblock Graph neural networks in network neuroscience.
\newblock arXiv preprint arXiv:2106.03535 (2021)

\bibitem{bessadok2020}
Bessadok, A., Mahjoub, M.A., Rekik, I.:
\newblock Topology-aware generative adversarial network for joint prediction of
  multiple brain graphs from a single brain graph.
\newblock International Conference on Medical Image Computing and
  Computer-Assisted Intervention (2020)  551--561

\bibitem{zhang2020b}
Zhang, L., Wang, L., Zhu, D.:
\newblock Recovering brain structural connectivity from functional connectivity
  via multi-gcn based generative adversarial network.
\newblock International Conference on Medical Image Computing and
  Computer-Assisted Intervention (2020)  53--61

\bibitem{isallari2020}
Isallari, M., Rekik, I.:
\newblock Gsr-net: Graph super-resolution network for predicting
  high-resolution from low-resolution functional brain connectomes.
\newblock International Workshop on Machine Learning in Medical Imaging (2020)
  139--149

\bibitem{pan2018}
Pan, S., Hu, R., Long, G., Jiang, J., Yao, L., Zhang, C.:
\newblock Adversarially regularized graph autoencoder for graph embedding.
\newblock arXiv preprint arXiv:1802.04407 (2018)

\bibitem{simonovsky2017}
Simonovsky, M., Komodakis, N.:
\newblock Dynamic edge-conditioned filters in convolutional neural networks on
  graphs.
\newblock Proceedings of the IEEE conference on computer vision and pattern
  recognition (2017)  3693--3702

\bibitem{zemouri2020}
Zemouri, R.:
\newblock Semi-supervised adversarial variational autoencoder.
\newblock Machine Learning and Knowledge Extraction \textbf{2} (2020)  361--378

\bibitem{goodfellow2014}
Goodfellow, I.J., Pouget-Abadie, J., Mirza, M., Xu, B., Warde-Farley, D.,
  Ozair, S., Courville, A., Bengio, Y.:
\newblock Generative adversarial networks.
\newblock arXiv preprint arXiv:1406.2661 (2014)

\bibitem{ronneberger2015}
Ronneberger, O., Fischer, P., Brox, T.:
\newblock U-net: Convolutional networks for biomedical image segmentation.
\newblock International Conference on Medical image computing and
  computer-assisted intervention (2015)  234--241

\bibitem{kipf2016}
Kipf, T.N., Welling, M.:
\newblock Semi-supervised classification with graph convolutional networks.
\newblock arXiv preprint arXiv:1609.02907 (2016)

\bibitem{gurler2020}
G{\"u}rler, Z., Nebli, A., Rekik, I.:
\newblock Foreseeing brain graph evolution over time using deep adversarial
  network normalizer.
\newblock International Workshop on PRedictive Intelligence In MEdicine (2020)
  111--122

\bibitem{fey2019}
Fey, M., Lenssen, J.E.:
\newblock Fast graph representation learning with pytorch geometric.
\newblock arXiv preprint arXiv:1903.02428 (2019)

\bibitem{fischl2012}
Fischl, B.:
\newblock Freesurfer.
\newblock Neuroimage \textbf{62} (2012)  774--781

\bibitem{dosenbach2010}
Dosenbach, N.U., Nardos, B., Cohen, A.L., Fair, D.A., Power, J.D., Church,
  J.A., Nelson, S.M., Wig, G.S., Vogel, A.C., Lessov-Schlaggar, C.N.,  et~al.:
\newblock Prediction of individual brain maturity using fmri.
\newblock Science \textbf{329} (2010)  1358--1361

\bibitem{shen2013}
Shen, X., Tokoglu, F., Papademetris, X., Constable, R.T.:
\newblock Groupwise whole-brain parcellation from resting-state fmri data for
  network node identification.
\newblock Neuroimage \textbf{82} (2013)  403--415

\bibitem{loshchilov2018}
Loshchilov, I., Hutter, F.:
\newblock Fixing weight decay regularization in adam.
\newblock (2018)

\bibitem{glasser2016}
Glasser, M.F., Coalson, T.S., Robinson, E.C., Hacker, C.D., Harwell, J.,
  Yacoub, E., Ugurbil, K., Andersson, J., Beckmann, C.F., Jenkinson, M.,
  et~al.:
\newblock A multi-modal parcellation of human cerebral cortex.
\newblock Nature \textbf{536} (2016)  171

\end{thebibliography}
\bibliographystyle{splncs}
\end{document}